\documentclass[fleqn,usenatbib]{mnras}

\usepackage{newtxtext,newtxmath}
\usepackage[T1]{fontenc}

\DeclareRobustCommand{\VAN}[3]{#2}
\let\VANthebibliography\thebibliography
\def\thebibliography{\DeclareRobustCommand{\VAN}[3]{##3}\VANthebibliography}

\usepackage{graphicx}	% Including figure files
\usepackage{amsmath}	% Advanced maths commands
\usepackage[version=3]{mhchem} % Package for chemical equation typesetting
\usepackage{gensymb}

\newcommand{\kcalMol}{kcal mol$^{-1}$}
\newcommand{\kcalMolK}{kcal mol$^{-1}$ K$^{-1}$}
\newcommand{\rrate}{mol min$^{-1}$ g$^{-1}$}
\newcommand{\gccc}{g cm$^{-3}$}

\newlength{\pointwidth}
\newcommand{\allende}{A}
\newcommand{\murchison}{B}
\newcommand{\kgfull}{C}
\newcommand{\kgmagn}{D}
\newcommand{\pyroxene}{E}
\newcommand{\serpentine}{F}
\newcommand{\fesreact}{J}
\newcommand{\ssblank}{G}
\newcommand{\sicblank}{H}
\newcommand{\kgblank}{I}

\newcommand{\fe}{Fe}

\newcommand{\cu}{Cu}
\newcommand{\cut}{Cu$^{2+}$}

\newcommand{\hydrogen}{H$_{2}$}

\newcommand{\water}{H$_{2}$O}

\newcommand{\sic}{SiC}

\newcommand{\co}{CO}
\newcommand{\cotw}{CO$_{2}$}

\newcommand{\htco}{H$^{13}$CO$^+$}

\newcommand{\methanol}{CH$_{3}$OH}

\newcommand{\acetaldehyde}{CH$_{3}$CHO}

\newcommand{\formamide}{NH$_{2}$CHO}

\newcommand{\nt}{N$_{2}$}

\newcommand{\nht}{NH$_{3}$}

\newcommand{\no}{NO}

\newcommand{\stm}{S$^{2-}$}

\newcommand{\hts}{H$_{2}$S}

\newcommand{\cusof}{Cu(SO$_{4}$)$_{2}$}

\newcommand{\fes}{FeS}
\newcommand{\covelite}{Cu$_{6}$S$_{6}$}

\newcommand{\fetot}{Fe$_{2}$O$_{3}$}

\newcommand{\cuoh}{Cu(OH)$_{2}$}

\title[Reactivity of chondritic meteorites]{Reactivity of chondritic meteorites under \hydrogen-rich atmospheres: Formation of \hts.}

\author[V. Cabedo et al.]{
V. Cabedo,$^{1}$\thanks{E-mail: v.cabedo@hw.ac.uk}
G. Pareras,$^{2}$
J. Allitt,$^{1}$
A. Rimola,$^{2}$
J. Llorca,$^{3}$
H. H. P. Yiu,$^{1}$
M. R. S. McCoustra$^{1}$
\\
$^{1}$Institute of Chemical Sciences, Heriot-Watt University, Edinburgh, EH14 4AS, Scotland.\\
$^{2}$Departament de Qu\'imica, Universitat Aut\`onoma de Barcelona, 08193 Bellaterra, Catalonia, Spain.\\
$^{3}$Institut de T\`ecniques Energ\`etiques, Departament d'Enginyeria Qu\'imica, Universitat Polit\`ecnica de Catalunya, Barcelona, Catalonia, Spain.
}

\date{Accepted XXX. Received YYY; in original form ZZZ}

\pubyear{2024}

\begin{document}
\label{firstpage}
\pagerange{\pageref{firstpage}--\pageref{lastpage}}
\maketitle

\begin{abstract}
Current models of chemical evolution during star and planetary formation rely on the presence of dust grains to act as a third body. However, they generally ignore the reactivity of the dust grains themselves. Dust grains present in the protoplanetary phase will evolve as the solar system forms and, after protoplanets have appeared, they will be constantly delivered to their surfaces in the form of large aggregates or meteorites. Chondritic meteorites are mostly unaltered samples of the dust present in the first stages of the Solar System formation, that still arrive nowadays to the surface of Earth and allow us to study the properties of the materials forming the early Solar System. These materials contain, amongst others, transition metals that can potentially act as catalysts, as well as other phases that can potentially react in different astrophysical conditions, such as \fes. In this work, we present the reactivity of chondritic meteorites under \hydrogen-rich atmospheres, particularly towards the reduction of \fes\ for the formation of \hts\ and metallic Fe during the early phases of the planetary formation. We present the obtained results on the reaction rates and the percentage of \fes\ available to react in the materials. Additionally, we include a computational study of the reaction mechanism and the energetics. Finally, we discuss the implications of an early formation of \hts\ in planetary surfaces.
\end{abstract}


%%%%%%%%%%%%%%%%%%%%%%%%%%%%%%%%%%%%%%%%%%%%%%%%%%

%%%%%%%%%%%%%%%%% BODY OF PAPER %%%%%%%%%%%%%%%%%%

%%%%%%%%%%%%%%%%%%%%%%%%%%%%%%%%%%%
%%	     Introduction			%%
%%%%%%%%%%%%%%%%%%%%%%%%%%%%%%%%%%%

\section{Introduction}

    Since the first detection of interstellar molecules \citep{McKellar1940}, it has been understood that the origin of chemical complexity on Earth lies beyond the surface of our planet. With the advancement of technology, the number of molecules detected in different astrophysical environments has risen to more than 250, plus their isotopologues \citep{Endres2016}, some of them, such as \htco, \methanol, \acetaldehyde\ or \formamide\ having an important relation to prebiotic chemistry (see \cite{Guelin2022} for a review). Interstellar complex organic molecules (iCOMs), i.e., molecules found in space with at least 6 atoms one of them being carbon, are considered to be the precursors for the formation of more complex systems that potentially gave rise to life, which were accreted on the surface of our planet during its formation or immediately after, and evolved there. Geochemical and additional synthetic processes that occurred during the first years of planetary evolution would have used those iCOMs as reactants to promote further complex chemistry, re-combining and forming interacting and self-replicating systems \citep{Walde2005, Schulze-Makuch2008, Nakashima2018}. Understanding the chemical complexity that brought life to Earth means studying where iCOMS formed, how were they delivered to Earth, and what was their role in the synthetic mechanisms towards the emergence of life that occurred during the infancy of planet Earth.

    Because in most astrophysical environments conditions are harsh, with very low gas densities and temperatures as low as 10 K, the current understanding is that most complex reactions occur in the surface of dust grains. At these very low temperatures, volatiles condense on the surface of dust grains, where molecules get closer together to form ices, and the likelihood for reactive encounters increases. Simple reactions, such as \co\ hydrogenation, can occur in this way and many observational and experimental works support this hypothesis \citep{Garrod2006, Garrod2008, Rimola2014, Linnartz2015, MartinDomenech2020, EnriqueRomero2022}. However, these models only reflect the role of the dust grain as a third body, increasing the density of the molecules on its surface, and absorbing the excess energy of the reaction to prevent the new molecule from dissociating \citep{Ferrero2023}. In these models, reactions are mostly governed by the interactions of molecules with the ice, and generally any reactivity from the grain itself is ignored. Hence, they are only valid in regions where dust grains are covered in ices, and do not take into account situations where the grains are bare \citep{Bennett2013, Marchione2019, Potapov2020} or consider warmer environments where ices might not present, such as very energetic regions of outflows or accretion or asteroidal and planetary surfaces.

    Chondritic meteorites are one of the closest samples that we can get nowadays from the dust that was present in the protoplanetary disk during the formation of the Solar System. Chondrites come from undifferentiated parent bodies, which were formed from the agglomeration of dust grains, and which experienced very low chemical or thermal alteration during their lifetime \citep{Zolensky2008}, hence preserving their pristine composition, allowing us to study the evolution of the primordial protoplanetary dust. Chondrites are agglomerations of different components ranging from submillimeter- to centimeter-size and which include: chondrules, Ca-Al-rich inclusions (CAIs), aggregates of olivines and other silicates, transition metal inclusions, such as Fe and Ni, metal oxides, such as \fetot, and sulfur-bearing compounds, such as \fes\ \citep{Weisberg2006}. Chondrites can be of three different classes, Enstatite (EC), Ordinary (OC) or Carbonaceous (CC), which differ in their refractory abundances with respect to solar abundances. Each class is, at the same time, sub-divided in different types, such as CMs or CVs (CCs), or H, L and LL (OCs). For a complete description of the classification of chondrites and their characteristics, see \cite{Weisberg2006}. Chondrites also comprise the largest proportion of meteorites falling to Earth surface, with CCs being particularly important, due to their high content in carbonaceous material.

    Recent studies suggest that the minerals present in meteorites and hence in dust grains, such as silicates, or other inclusions, such as transition metals or metallic oxides, can have proper catalytic activity, i.e., can lower activation energies and/or promote new reaction mechanisms when interacting with the gas and/or the liquid phase on their surface \citep{Llorca1998, Llorca2000, Ferrante2000, Kress2001, Rotelli2016, Cabedo2021a, Peters2023, Pareras2023, Pareras2024}. The material present in the solid phase during the star formation process would have been inherited and delivered to accretioning protoplanets, and would have enhanced the chemical complexity of the forming planets, potentially promoting chemistry which is key to the origin of life \citep{Birnstiel2016, Nakashima2018}. In this context, meteoritic material can be of key importance for the delivery of solids, water, and already formed complex molecules, but also for the chemical evolution on the early Earth surface, and hence for pre-biotic chemistry. 

    The metallic elements present on chondrites are well known on Earth for their catalytic properties, which are widely used in the industry. For example, \fe\ and iron oxides catalysts are extensively used for processes such as the production of fuels through the Fischer-Tropsch (FT) process \citep{Mahmoudi2017}, or the production of \nht\ through the Haber-Bosch (HB) mechanism \citep{Liu2014}. Previously, some of the present authors presented an experimental study showing the reactivity of chondritic material, and in particular their metallic inclusions, towards the formation of small hydrocarbons following the FT mechanism \citep{Cabedo2021a}. Additionally, we have performed computational analysis that point towards the potential ability of single atom particles present in dust grains to carry out the formation of small hydrocarbons and alcohols through the FT \citep{Pareras2023, Pareras2024}. Here, we aimed at extending both the experimental and computational work towards the direct hydrogenation of chondritic material with \hydrogen, under a \nt\ atmosphere at appropriate protoplanetary conditions. In particular, we aim at proving the hydrogenation of \fes\ towards the production of \hts\ and metallic Fe, following: 

    \begin{equation}
        \ce{FeS}{(s)} + \ce{H2}{(g)} \rightleftharpoons \ce{Fe}^{0}{(s)} + \ce{H2S}{(g)}.
        \label{eq:h2s_formation}
    \end{equation}
    
    We follow the same experimental procedure as in \cite{Cabedo2021a}, but we change the analysis system for a \cusof\ solution (see the Section \ref{sec:exp_proc}), expecting that the formed \hts\ will dissolve in water, and \stm ions will quickly react with \cut\ and precipitate in the form of \covelite. We complement the experimental study with a computational evaluation of the viability of the mechanism at the experimental conditions. We present here the results and discuss its implication towards chemical evolution in different astrophysical environments, in particular on the primitive Earth surface.

%%%%%%%%%%%%%%%%%%%%%%%%%%%%%%%%%%%
%	Experimental Procedure			%
%%%%%%%%%%%%%%%%%%%%%%%%%%%%%%%%%%%    

\section{Experimental procedure} \label{sec:exp_proc}

    \subsection{Sample description and preparation} \label{sec:sampleDescription}

        Different samples were used to check the reactivity under the given conditions. Meteoritic samples are chosen to prove their surface reactivity. They are provided by different collections and used according to their availability. CCs and OCs are chosen because of their primitive composition, and their relative low degree of alteration and lack of phase differentiation. CCs are characterised by having Mg/Si ratios near solar value and O isotopic composition near or below the terrestrial fractionation line while OCs are distinguished by presenting sub-solar Mg/Si and refractory/Si ratios, to exhibit O isotope composition above the terrestrial fractionation line. These characteristics make them good representatives of the composition of the dust present in the protoplanetary disks and the surfaces of protoplanets during accretion \citep{Weisberg2006}. Both classes are also representative of the largest numbers of meteoritic falls on Earth surface and, by extrapolation, of the material that fell on Earth during the first stages of planetary evolution. 

        Allende is the largest CC to have fall on Earth, emitting an impressing fireball, in the early morning of 1969, in Pueblito de Allende, Mexico. Numerous specimens of this meteorite have been recovered during the years, making it one of the most studied meteorites. Allende is classified as a CV which are characterised by having an abundant matrix (up to 40 vol\%), large chondrules (mm sized) many of which are surrounded by igneous rims, and a high abundance of Calcium-Aluminium-Inclusions (CAIs). Allende is an oxidised CV meaning that it has a coarser matrix and is more extensively altered than its reduced counterparts. All CVs are of petrologic type 3, which have abundant chondrules, present low degrees of aqueous alteration and have unequilibrated mineral distributions.

        Murchison fell in 1969 near Murchison, Australia. Due to its large mass it is also a very well studied meteorite. It became very well known due to the large diversity of organic compounds detected on its interior \citep{ScmittKopplin2009}. Classified as a CM, Murchison is one of the least-altered chondrites. CMs are distinguished by small chondrules and refractory inclusions (0.3 mm), abundant matrix (around 70 \%), and abundant hydrated minerals such as phyllosilicates. Murchison is of petrologic type 2, where only some sillicates have been aqueously altered. CM-like material usually occurs in clasts in other chondrite groups and achondrites, suggesting a wide distribution of the material in the early solar system \citep{Zolensky1996, Weisberg2006}. Both Allende (Sample \allende) and Murchison (Sample \murchison) were obtained as fine ground powders. 
        
        KG 007 is an OC of subtype H and petrological type 6. OCs contain a large volume percentage of chondrules, with only 10-15 vol\% of fine-grained matrix. They do not contain organic matter in their matrix. The H (high-iron) group of OCs are characterised by their high siderophile element content, with metal abundance of around 8 vol\%. Petrological type 6 designates chondrites that have been metamorphosed under conditions sufficient to homogenise all mineral compositions, but melting did not occur. KG 007 was obtained fully ground and in two different phases: one containing the full meteoritic composition (Sample \kgfull) and another one containing only metallic inclusions separated by magnetic means (Sample \kgmagn).
        
        Silicate samples of serpentine and pyroxene, which are minerals commonly found in meteorites \citep{Weisberg2006}, are used to asses where the reactivity is originating and if it is due to the presence of silicates, the potential catalytic ability of the metallic inclusions or other type of reactivity. 
        
        Additionally, we made three different blank experiments: one only with the stainless steel (SS)  reactor (Sample \ssblank), one with only SiC (Sample \sicblank) and one containing a sample of KG 007 (full composition) and 40 ml/min of \nt\ (Sample \kgblank) to prove that products are not produced by the out-gassing of the meteorite. Finally, the \fes\ experiments (Sample J-J.5) are carried out with commercial \fes\ from Thermo Fischer Scientific $\copyright$, with 99.9 \% purity, and a mesh size of 100 $\mu$m.
        
        Preparation of the samples prior to introduction on the reactor is explained in our previous work \citep{Cabedo2021a}. 

    \subsection{Experimental set-up for measuring the samples reactivity} \label{sec:setup}

        The total weight of the samples is introduced in a stainless steel (SS) reactor heated by a furnace where a flux of \hydrogen\ and \nt\ are circulated. Gaseous products are circulated through a quartz bubbler and collected in a 0.1 M \cusof \ solution. We expect the production of \hts\ following Eq. \ref{eq:h2s_formation}, and the subsequent precipitation of \covelite. \cu\ is chosen for its ability to react, the high constant of the reaction for the solid precipitation, and the easiness of qualitatively identifying the production of \covelite\ as a black solid. Additionally, it allows its quantification by collecting the resulting \covelite. Any solid left on the bubbler is scraped and left in the solution, which is transferred to a conic bottom flask and left to deposit the solid. The liquid is then decanted, and the solid is rinsed with water and decanted again, three times. The solid is extracted from the flask with acetone, and left to dry for 48 h at 80 \degree C. Finally, the solid is weighed. The error in this procedure can be high and unknown, due to the difficulty of removing all the solid precipitated inside of the bubbler, however, it makes a qualitative assessment extremely simple and provides a lower-limit for the reaction considered.
        
        Reaction conditions are chosen to resemble those present in different warm astrophysical environments, such as dust grains in solar nebulae experiencing high temperature conditions (shocks, accretion streams), or the surface of rocky protoplanets, like Earth.

        \begin{itemize}
        
            \item The ratio of reactants was set to \hydrogen:\nt\ = 3, as a likely ratio present in the solar nebula or a primitive Earth atmosphere \citep{Zahnle2010, Wordsworth2016}. The fluxes of the reactants were set to a total of 40 mL/min, with \hydrogen\ $\approx$ 32 mL/min and \nt\ $\approx$ 8 $\pm$ 1 mL/min.
            
            \item The temperature is set at 873 K. This is a relevant temperature for the type of reactions studied, which can be achieved in the astrophysical environments proposed.

            \item Reactions are left to proceed for 48 h, in order to accumulate as much solid product as practically possible.
            
        \end{itemize}

    \subsection{Characterisation of solid products by X-Ray diffraction} \label{sec:xray_method}

        To analyse the samples through X-Rar Difraction (XRD), a small amount (few mg) of the dried solid produced during the reaction is taken to an X-Ray diffractometer in order to asses its composition and cristalographic structure. The powder is dispersed in isopropyl alcohol on a silicon zero background holder. Data were collected on a Malvern Panalytical MultiCore Empyrean diffractometer with generator settings of 40 mA and 45 kV. The sample holder was placed  on the Reflection-Transmission Spinner stage  and the sample was rotated at 15 rpm. Intensity data were collected in Gonio mode (Bragg Bretano geometry). Data were collected in the range of 5-85 in 2$\Theta$. The Step size was  0.0260 \degree\ and the scan time was around 30 mins. The diffraction pattern was analysed using the HighScore software package from Malvern Panalytical. Automatated background subtraction, peak identification and searching of the Powder Crystallographic open database was performed with pre‑programmed batch ``IdeALL'' provided in HighScore. Fit of likely patterns was checked with basic Rietveld refinement using the MultiRiet script and the Chi Square, R$_{p}$, R$_{wp}$ and R$_{exp}$ values.

    \subsection{Magnetic susceptibility analysis} \label{sec:magneticsusc_method}

        The magnetic susceptibility, $\chi$, is the ability of a magnetic material to react to an external magnetic field. Following:

        \begin{equation}
            M = \chi\ H
            \label{eq:magnetic_susceptibility}
        \end{equation}

        it relates a material's magnetisation, $M$, with the strength of an applied magnetic field, $H$. Conventionally, these terms are assumed to be linearly dependent, which is true for high temperatures and low fields. Typically, magnetic susceptibility has been measured by determining the apparent weight of a sample under a magnetic field, but more precise and simpler techniques have been developed in the recent years that allow for a quick characterisation of the magnetic properties of samples. For a complete description on magnetic susceptibility see \cite{Mugiraneza2022}. In here, it suffices to say that we are interested in detecting any change in the susceptibility of the samples before and after the reaction, which would implicate a change in the magnetic composition, and potentially, the metallic one. Since our samples are very heterogeneous, we define a total specific susceptibility, $\chi_{s}$, for the complete sample by measuring the given susceptibility, $\chi$, and dividing by the total mass of each sample, $m$:

        \begin{equation}
            \chi_{s} = \chi / m
            \label{eq:specific_susceptibility}
        \end{equation}

        Measurements of the magnetic susceptibility of the samples are done with a magnetic susceptibility meter from ZH Instruments. The measurements are done by introducing a representative amount of each sample in a plastic bag and shaped in a layer of around 5mm of thickness. Measurements are taken directly next to the samples with no air gap between the instrument and the samples, so no thickness correction needs to be applied. We measure the magnetic susceptibility before the reaction, $\chi_{s}^{0}$, by taking a representative amount of the samples without reacting (i.e. the meteoritic material), and the magnetic susceptibility after reaction, $\chi_{s}^{r}$, by taking the whole amount of the diluted samples once reacted.

    \subsection{Computational details} \label{sec:compu_details}

        The reactive surfaces were modelled using a periodic approach. Quantum chemical calculations were performed with the CP2K package \citep{Kuhne2020}. The characterization of the potential energy surface (PES) of the reactions requires determining the structures and the energetics of the stationary points. For geometry optimizations, the semi-local PBE$_{sol}$ functional was used \citep{Perdew2008}, along with the Grimme D3(BJ) correction to include dispersion forces \citep{Grimme2010}. A double-$\zeta$ basis set (DZVP-MOLOPT-SR-GTH gaussian basis set) was adopted for all the atom types, combined with a cut-off set at 500 Ry for the plane wave auxiliary basis set \citep{VandeVondele2005, Kuhne2020}. The Goedecker–Teter–Hutter pseudopotentials were used to describe core electrons, while a mixed Gaussian and plane-wave (GPW) approach was employed for valence electrons \citep{Goedecker1996}.

        The bulk of troilite (stoichiometric \fes) exhibits an hexagonal structure (space group P$\bar6$2c ), with XRD crystallographic cell parameters of a = b = 5.896 (\r{A}) and c = 11.421 (\r{A}), and $\alpha$ = $\beta$ = 90 degrees and $\gamma$ = 120 degrees, which upon PBE$_{sol}$ optimization convert into a = 5.697 b = 5.580 (\r{A}) and c = 10.208 (\r{A}), and $\alpha$ = $\beta$ = 90 degrees and $\gamma$ = 120 degrees. A recent study considering different slabs generated automatically by the code POLYCLEAVER \citep{MatesTorres2024}, concluded that the most stable surface is 011 \citep{MartinezBach2024}, and so we choose that surface for all our studies. To approach our system to reality we generated different defects by means of Fe vacancies. The \fes(011) structure shows a combination of tetra- and penta-coordinated Fe atoms, to generate the defective structures rationally, we have constructed geometries in which the vacancies are from a tetra-coordinated Fe (Tetra-Def) or a penta-coordinated (Penta-Def). Moreover, we have also considered the combination of multiple defects on the same structure, removing in this case two tetra- or penta-coordinated irons (Tetra-Def-2 and Penta-Def-2 respectively) or one of each Fe (Tetra-Penta-Def). Note that to maintain the stoichiometry (and hence the electroneutrality), the relevant S atoms were removed from the opposite face of the slab.

        The climbing image nudged elastic band (CI-NEB) technique implemented in CP2K \citep{Kuhne2020} was used to search for transition states. The CI-NEB calculations were also run at PBEsol-D3(BJ) level. Activation energy barriers were calculated as
    
        \begin{equation}
            \Delta E^\ddagger = E_{TS} - E_{GS},
        \end{equation}
        \begin{equation}
            \Delta U^\ddagger = \Delta E^\ddagger + \Delta ZPE,
        \end{equation}
        \begin{equation}
            \Delta G^\ddagger _T = \Delta E^\ddagger + \Delta G_T,
        \end{equation}
        
        where $\Delta$E${^\ddagger}$ is the potential energy barrier, in which E$_{TS}$ and E$_{GS}$ are the absolute potential energies for the transition state and the local minimum structure, respectively, $\Delta$U${^\ddagger}$ is the vibrational zero-point energy (ZPE) corrected barrier, in which $\Delta$ZPE refers to the contribution of the ZPE corrections to $\Delta$E${^\ddagger}$, and $\Delta$G$_T^\ddagger$ the Gibbs activation barrier at a given temperature, in which $\Delta$G$_T$ refers to the contribution of the Gibbs corrections to $\Delta$E${^\ddagger}$. Note that final Gibbs energies have been computed at the working temperature of 800 K. 
        
        The nature of the stationary points of the reactions was validated by calculating the harmonic frequencies; the outcome being minima for reactants, intermediates and products, and first-order saddle points for transitions states. Vibrational harmonic frequencies were calculated at the PBE$_{sol}$-D3BJ/DZVP-optimized structures using the finite differences method as implemented in the CP2K code \citep{Kuhne2020}. To minimize the computational cost, a partial Hessian approach was employed. Consequently, vibrational frequencies were computed for a subset of the entire system, comprising only the surface atoms participating in the reaction and the reactive species.

%%%%%%%%%%%%%%%%%%%%%%%%%%%%%%%%%%%%%%%%
%	Results   %
%%%%%%%%%%%%%%%%%%%%%%%%%%%%%%%%%%%%%%%%

\section{Results} \label{sec:results}

    \subsection{Reactivity of the chondritic meteorites towards the formation of \hts} \label{sec:h2s_formation}

         In this section we describe the results regarding the reactivity of the chondritic meteorites under reducing conditions. The different samples, and the results for all the experiments are summarised in Table \ref{table:sampleDescrp}. First, we carried out the blank experiments (samples \ssblank, \sicblank\ and \kgblank). No precipitate is formed so we can confirm that i) there is no activity coming from the SS reactor or from the \sic\ used as solvent, and ii) no product formed is the result of out-gassing of the samples. 
    
        \begin{table*}[!ht]
                \centering
                \caption{Sample description and results}
                \begin{tabular}{c c c c c c c c c}
                \hline
                \hline
                \vspace{-5pt}
                &\\
                 & Sample & Weight & \covelite weight & \fes & \fes & $\chi_{s}^{0}$ & $\chi_{s}^{r}$ & $\Delta\chi_{s}$ \\
                & & (g) & (g) & (\%wt) & (\%vol) & (g$^{-1}$) & (g$^{-1}$) & (g$^{-1}$) \\
                \hline
                &\\
                A & Allende & 0.5002  & 0.0211 & 17.3  & 10.48 & 0.1331 & 1.353 & 1.220\\
                B & Murchison & 0.5000 & 0.0200 & 16.4 & 8.52 & 0.2249 & 3.250 & 3.025 \\
                C & KG007 & 0.5000 & 0.0045 & 3.5 & 2.69 & 0.7616 & 3.710 & 2.948 \\
                D & KG007 & 0.5001 & 0.0057 & 4.7 & - & 1.333 & 4.539 & 3.205\\
                 & (magnetic phase) & & & & &  \\
                E & Pyroxene & 0.5003 & $<$ 0.0001 & $<$ 0.1 & - & 0.1706 & 0.6646 & 0.4939 \\
                F & Serpentine & 0.5002 & 0.0007 & 0.6 & - & 0.0375 & 0.4368 & 0.3992 \\
                G & SS* blank & - & - & - & - & - & - & - \\
                H & SiC blank & 0 & 0 & - & - & - & - & - \\
                I & KG007 blank & 0.5004 & 0 & - & - & 0.7616 & 2.188 & 1.426 \\
                J & FeS & 0.5002 & 0.0423 & - & - & 0.0148 & 1.9462 & 1.931 \\
                \hline
                \end{tabular}
                    \begin{list}{}{}
                        \item Instrumental errors are $\pm$ 0.0001 g. Errors for the percentages on weight and volume are derived from instrumental errors and are of the order of $\pm$ 0.1.
                        \item * Stainless steel (SS)
                    \end{list}
                    \label{table:sampleDescrp}
                \end{table*}
    
\begin{figure}
                \includegraphics[width=\columnwidth]{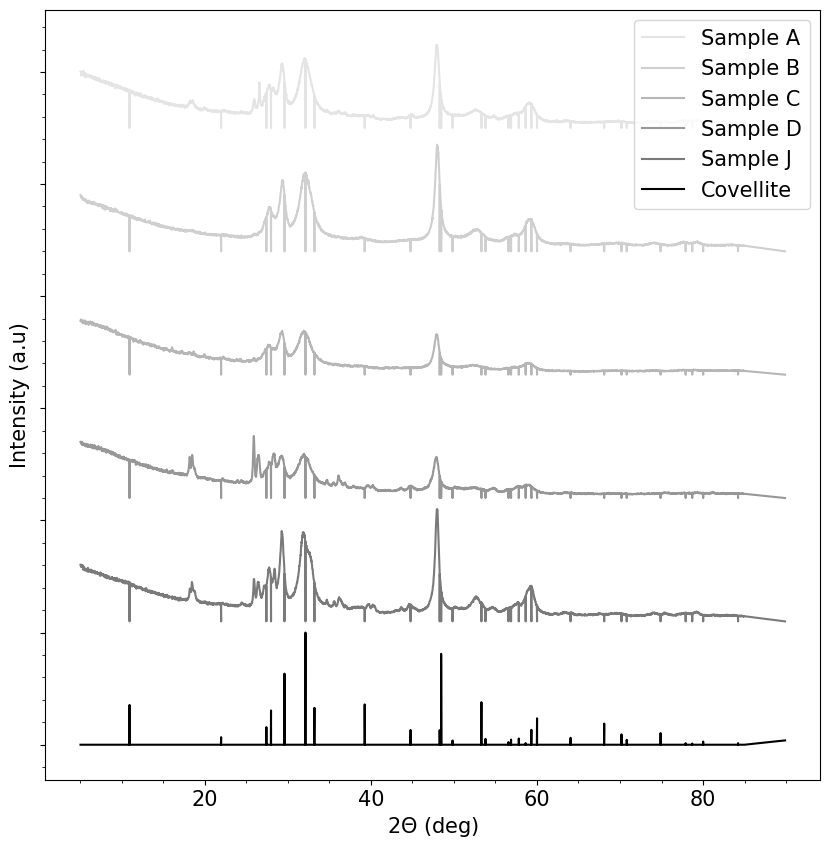} \\
            \caption{X-Ray Diffraction patterns of Covellite (bottom) and the recovered products. The downward lines in the spectra correspond to the theoretical position of the covellite peaks. Note that Samples \pyroxene\ and \serpentine\ are not shown due to a lack of sample to perform the XRD analysis.}
            \label{fig:xray_covellite}
        \end{figure}

         For all meteoritic samples (\allende, \murchison, \kgfull\ and \kgmagn), a black solid is observed precipitating in the solution soon after reaction starts. The same black precipitate is observed when reacting the silicate samples (Samples \pyroxene\ and \serpentine). In the case of pyroxene (Sample \pyroxene) the amount of solid formed was so small that it was not possible to retrieve it from the solution. XRD patterns of the recovered solid are shown in Fig. \ref{fig:xray_covellite}, which shows that the solid obtained is \covelite\ (Covellite) in all cases. The experimental spectrum of covellite is shown for comparison \citep{Takeuchi1985, AMCSD}. The downward lines in the experimental spectra correspond to the theoretical position of the covellite peaks. Additional peaks correspond to Poitevinite, a phase of \cusof\ left as a residue from the used solution. In order to confirm that the produced \covelite\ is due to the reduction of \fes, we perform the same experiment with commercial \fes\ as reactant (Sample \fesreact). The production of \covelite\ in the solution was also observed, confirmed by the XRD pattern (Fig. \ref{fig:xray_covellite}), which allows us to confirm that a mechanism similar to Eq. \ref{eq:h2s_formation} is happening.
    
        The reaction of the sulfides present in the samples with \hydrogen\ should increase the magnetic properties of the sample for two reasons: i) the removal of Fe and S atoms from the troilite (\fes) lattice and the subsequent formation of pyrrothites (Fe$_{1-x}$S) of unknown stoichiometry increases the magnetisation of the samples since the former is non-magnetic but the later is \citep{Dorogina2015}, and ii) the formation of metallic Fe and iron oxide nanoparticles with strong magnetic properties \citep{Woo2004}. Hence, we measured the magnetic susceptibility before ($\chi_{s}^{0}$) and after ($\chi_{s}^{r}$) reaction. The results of these measurements is shown in Table \ref{table:sampleDescrp}. There is a clear increase in the $\chi_{s}$ values before and after reaction, suggesting an increase on the magnetic components of the sample, and further confirming that the mechanism proposed is taking place. The change in the magnetism of \fes\ is evident in Sample \fesreact\, where the only phase present initially is troilite. However, the obtained values suggest that additional reactions are taking place which cannot be quantified. We discuss this further in Section \ref{sec:general_reactivity}. 
    
        We can estimate the amount of \fes\ in our samples by deriving  the reaction rate, $k$, for Eq. \ref{eq:h2s_formation} from the commercial sample by making certain assumptions. First we assume that the composition of the commercial \fes\ sample is uniform, and that only 69.2 $\%$ is \fes, all in the form of troilite, which reacts with \hydrogen. Secondly, we assume that \hydrogen\ has a large enough contact time with the sample to react completely, and that the contact is isotropic and equal through all the reactor. Finally, we assume that the reaction rate is constant through all the reaction time, as there is enough of both reactants. We can derive $k$ such as:
    
            \begin{equation}
                k = \frac{n_{H_2S}^{f}}{g_{FeS} t}
            \end{equation}
    
        where $n_{H_2S}^{f}$ are the number of mols of \hts\ that have been produced, $g_{FeS}$ is the initial mass of \fes\ in the commercial sample, in grams, and $t$ is the reaction time (which in the case of Sample \fesreact\ is 24 h). We obtain a value of 8.877 $\pm$ 0.002 $\times$ 10$^{-7}$ \rrate\ at 873 K. With this value in hand, and assuming that all the different phases of \fes\ present in the samples react with the same rate constant as troilite, and that the grain size is similar through all the samples and its influence on the reaction rate is negligible, we can approximate the percentage in weight (\%wt) of \fes\ in the totality of the samples. To obtain the percentage in volume (\%vol) we use bulk density values for Samples \allende, \murchison\ and \kgfull\ of 2.8, 2.4 and 3.7 \gccc\ respectively \citep{Flynn1999}, and a density value of 4.6 \gccc\ for Sample \fesreact\ \citep{RUFFdatabase}. 
    
        In the literature, the sulfide content in chondrites varies depending on the type and alteration history of the parent body to which the sample belongs to.  For CMs, which have suffered a certain degree of aqueous alteration \citep{Brearley2006, Singerling2018}, the \%vol ranges from 0.6 to 5.9, with 1.8 \%vol for Murchison \citep{Howard2010, Howard2015, Schrader2021}. We obtain a larger value of 8.52 \% vol. Less altered CVs show larger ranges and higher values, from 1.9 to 9.9 \%vol, with 6.6 \%vol for Allende \citep{Howard2010, Schrader2021}. OCs of type H show various degrees of alterations, H6 are highly thermally and aqueously altered and show more restricted values and higher, from 4.1 to 7.0 \citep{Schrader2021, Dunn2010}. Our values seem to be slightly too high for CCs but slightly too low for the OCs. We note that the scope of this work was not to accurately determine the amount of sulfides in chondritic meteorites, but to prove their reactivity, and so we consider that our results are close enough to values in the literature to consider our hypothesis valid. However, we note here some reasons why our values might be slightly inaccurate. First, we note that in the literature the most common technique to determine the composition of a sample is XRD, although other techniques, such as Electron Microscopy (EM) can be used. XRD is considered accurate up to $\pm$ 2 \%, as long as the sample that is being analysed is considered representative of the whole sample, and the fitting of the spectral matching is good \citep{Bland2004}. On the other hand, our method contains many approximations, which have been mentioned through the text, and is a simplification of what is possibly a complicated network of chemical reactions and changes in the solid phase. Particularly important are the assumptions on the reaction rates and the densities used to obtain the \%vol. Additionally, our ability to replicate the experiments in order to obtain independent measures is completely restricted by the amount of samples that we can analyse. Finally, we note that we cannot recover all the \covelite\ from the solution, and so we can only present a minimum of the \fes\ measured through this method. 
        
        Nevertheless, we can make a rough analysis of the relation between the alteration degree, and the amounts of \fes\ found in our sample. Sulfur containing minerals are believed to condense from the solar nebula mostly in the stoichiometric form of troilite, and is later processed through aqueous and thermal alteration, becoming an indicator of the alteration degree of the parent object \citep{Zolensky1995, Kimura2011, Schrader2019, Schrader2021}. We find that the amount of product, and hence the amount of \fes\ found in the sample correlates well with the degree of alteration, as we find a decreasing amount of \fes\ (OC $\leq$ CM $\leq$ CV) with increasing alteration degree (OC $\geq$ CM $\geq$ CV). Additionally, this points to the fact that the phase more susceptible to hydrogenation is troilite. Sulfurisation processes, i.e., the formation of troilite phases from \hts\ and \fe\ \citep{Lauretta1998}, could be occurring once \hts\ is formed in the reactor, however, we have no evidence of that and instead our results point towards the reverse process being more favourable at the reaction conditions.  
    
        Finally, we can estimate the activation energy of the reaction, $E_{a}$, by performing additional experiments with commercial \fes\ at temperatures from 323 to 873 K. We observe that at 573 K some \covelite\ is produced, however, the amount is minimum and cannot be recovered. Below that temperature, no \covelite\ production is observed and no results are presented. Details of these experiments are shown in Table \ref{table:fes_experiments}.
    
            \begin{table*}
                \centering
                \caption{Results of the \fes\ experiments}
                \begin{tabular}{c c c c c}
                    \hline
                    \hline
                    \vspace{-5pt}
                    &\\
                    Sample & Temperature & \fes\ weight &  \covelite\ weight & k \\
                    & (K) & (g) & (g) & ($\times$10$^{-7}$\rrate) \\
                    \hline
                    &\\
                    J & 873 & 0.5002 & 0.0423 & 8.877 \\
                    J.1 & 823 & 0.5003 & 0.0349 & 7.322 \\
                    J.2 & 773 & 0.5002 & 0.0260 &  5.456\\
                    J.3 & 723 & 0.5004 & 0.0256 & 5.37 \\
                    J.3 & 673 & 0.5001 & 0.0068 & 1.477\\
                    J.4 & 623 & 0.4999 & 0.006 & 1.26 \\
                    J.5 & 573 & 0.5000 & $<$ 0.0001 & $<$ 1\\
                    &\\
                    \hline
                \end{tabular}
                \begin{list}{}{}
                        \item Instrumental errors are $\pm$ 0.0001 g. Errors in the reaction rate  are derived from instrumental errors and are of the order of $\pm$ 0.02 $\times$ 10$^{-7}$\rrate.
                    \end{list}
                \label{table:fes_experiments}
            \end{table*}
    
            We derive the activation energy of the reaction from the Arrhenius equation:
    
            \begin{equation}
                k = A \exp{\frac{-E_{a}}{k_{B}T}}
                \label{eq:arrhenius}
            \end{equation}
    
            where $A$ is the pre-exponential factor (in \rrate), $k_{B}$ is the Boltzmann constant (in \kcalMolK) and $E_{a}$ is the activation energy (in \kcalMol). We find that the reaction constant behaves linearly between 573 and 723 K, and then flattens. Following Eq. \ref{eq:arrhenius}, we find $E_{a}$ = 44.4 \kcalMol. We discuss this value along with the computational results in Section \ref{sec:hts_desorption}.

    \subsection{Further reactivity of the samples} \label{sec:general_reactivity}

        When looking at the magnetic susceptibility results, we observe interesting features that point to additional reactivity not related to the production of \hts. First, the increase in the magnetic susceptibility of Sample \kgblank\ is striking. Since \hydrogen\ is not present in this experiment we attribute the increase to the thermal decomposition and phase changes of different compounds in the sample, such as \fes\ (pyrrothites, but also troilite) \citep{Suttle2021} and Fe oxides \citep{Jozwiak2017}. Additionally, \nt\ could participate in reactions with Fe compounds that we cannot detect, but this scenario is unlikely. Although NS, and related molecules, are widely observed in different regions of space \citep{McGonagle1994,McGonagle1997, Sanz-Novo2024} their formation in the experimental conditions is complicated due to the low reducing capacity of \nt\ and an unfavourable mechanism \citep{Pereira2010}. Abundance of nitrogen oxides, such as \no, is also large in the ISM \citep{Dupuy2017}, but the direct reduction of Fe oxides with \nt\ seems unlikely. We note as well, that no production of \nht\ is observed through the HB mechanism, which would be seen by the precipitation of \cuoh, a light blue solid, in the solution.
    
        Secondly, we note that the increase in magnetic susceptibility is not directly correlated with the amount of \fes\ found for each sample, additionally suggesting that other reactions are taking place. In addition to all the possible reactions already mentioned, iron oxides, largely present in all of the samples, can be  reduced in the presence of \hydrogen\ to form other oxides, metallic iron, and water, at the reaction temperature \citep{Jozwiak2017}, contributing to the increase of magnetic susceptibility by producing metallic \fe. Reduction of iron silicates can also be occurring \citep{Massieon1993} to give metallic Fe. This process can be hinted at with Samples \pyroxene\ and \serpentine\, which contain very low amounts of \fes\ and have a very low initial magnetic susceptibility, which increases substantially after the reaction has taken place.
    
        Finally, we need to consider that our samples contain different phases of sulfides: the most abundant are of the pyrrhotite group (Fe$_{1-x}$S where 0 $\leq$ x $\leq$ 0.125, but can be up to 0.2), In addition, it can also occur in forms of of pyrite (FeS$_{2}$) and pentlandine ((Ni,Fe)$_{x}$S$_{x-1}$) inclusions \citep{Bullock2005, Singerling2018, Schrader2021}. In this work we are only considering the reduction of pure troilite, but reduction of other phases would also result in the formation of \hts\ and metallic Fe, as well as an increase on the magnetic susceptibility.
        
        While we acknowledge that we cannot directly correlate the change in magnetic susceptibility to a particular reaction, we consider that it is evidence enough of the general reduction of the material. Since we cannot measure any of the additional products of the reduction with the simple experimental procedure, and the different samples contain different amounts and forms of iron oxides and sillicates, we cannot derive any reaction constants or provide any further characterisation. Further experiments with a more detailed separation and characterisation of the reaction products are needed.

    \subsection{Computational characterisation of the reaction mechanism} \label{sec:dft_mechanism}
    
        To complement our study, we provide density functional theory (DFT) calculations to characterise the mechanisms and derive the energetics of the \hts\ formation following Eq. \ref{eq:h2s_formation}. Although the nature of the \fes\ in the meteoric samples is diverse, we select a pure troilite structure to perform the calculations, following the procedure described in Section \ref{sec:compu_details}.

        \begin{figure*}
            \centering
            \includegraphics[width=\textwidth]{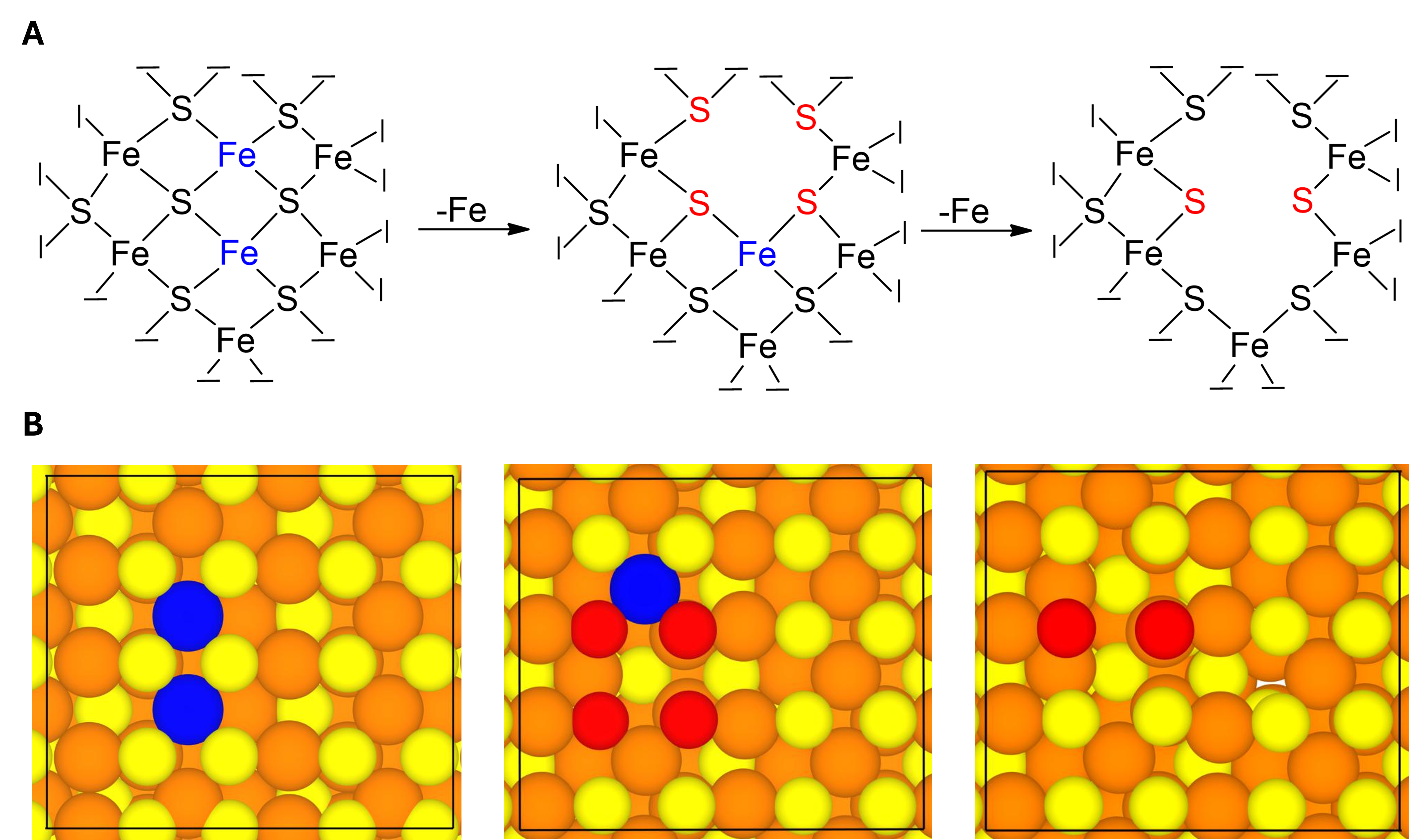}
            \caption{Panel A: Schematic representation for the formation of the Fe vacanciess on the (011)\fes\ surface. Panel B: optimized geometries for the original, non-defective, slab, the slab with a single Fe vacancy and the slab with two Fe vacancies (S atoms are in yellow and Fe atoms in orange). Fe atoms highlighted in blue colour are the ones removed on the process. S atoms highlighted in red are the most reactive ones in each structure.}
            \label{fig:geom_fe_vacancies}
        \end{figure*}
    
\begin{figure*}
        \centering
        \includegraphics[width=\textwidth]{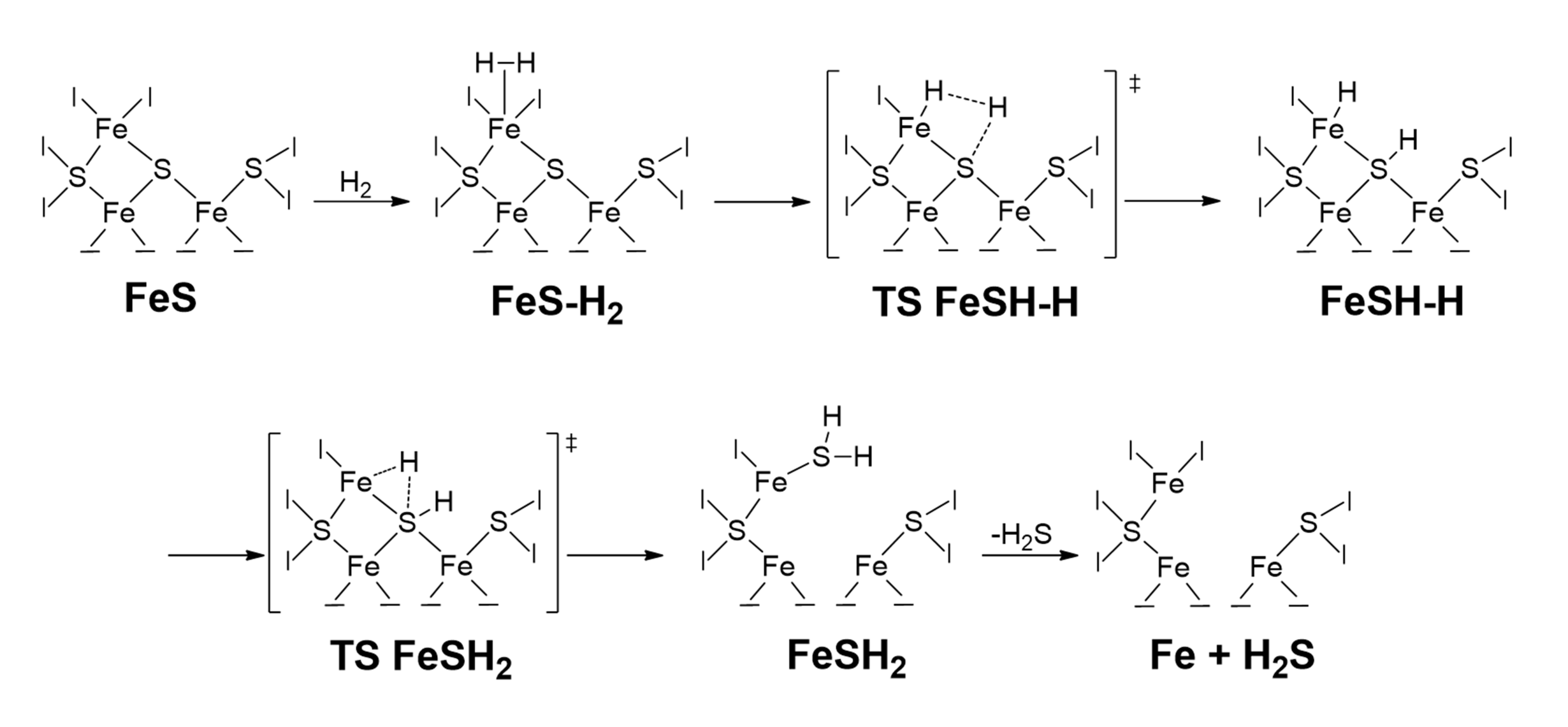}
        \caption{Proposed reaction mechanism for the \hts\ formation}
        \label{fig:compu_mechanism}
        \end{figure*}

        The reaction mechanism that we propose based on Eq. \ref{eq:h2s_formation} proceeds through four steps (see Fig. \ref{fig:compu_mechanism}): first, the adsorption of \hydrogen, second and third, the successive H transfers from the Fe to the S, and finally the desorption of \hts. We have constructed the PES for each of the proposed defective surfaces. In this section, we comment the obtained energies and the viability of the mechanism. A complete summary of the values can be found in the Supplementary Information.
    
        The \hydrogen\ adsorption can only happen on two possible positions: on the penta-coordinated or on the tetra-coordinated Fe surface atoms. The later showing larger stabilisation energies upon \hydrogen\ adsorption, because \hydrogen\ breaks in a homolytic fashion. Once \hydrogen\ has been adsorbed, the mechanism proposed proceeds through two hydrogen transfers from the Fe to the S, each of them with an associated energetic barrier. The values for the barriers and the reactions of the different proposed systems are shown in Table \ref{tab:GibbsE}. If we consider the systems where \hydrogen\ has been absorbed onto a penta-coordinated Fe, for the first hydrogenation we see values between 16 \kcalMol and 23 \kcalMol, and, in most of the cases, this step is endergonic. The second hydrogenation has slightly higher energy barriers. In all cases we see a reduction in the barrier when increasing the number of defects. The reported Gibbs energies show that the final hydrogenation is endergonic. For systems where \hydrogen\ is absorbed in the tetra-coordinated Fe, the \hydrogen\ adsorption is slightly more favourable. The first hydrogenation also proceeds with slightly lower energy barriers, reaching its lowest at 1 \kcalMol. This process is also exergonic for the more reactive configurations. Finally, barriers for the second hydrogen transfer show to be substantially higher, due to the over stabilisation of the remaining H on the surface. Calculated vales for the Gibbs energies for the last step also show that all the processes are endergonic. In summary, differences are observed between the different systems, however, although we treat the results as individuals one needs to consider that the real picture of the process will be a combination of all of them.

        \begin{table*}
            \centering
            \caption{Collected Gibbs-corrected energy barriers at 800K ($\Delta$G${^\ddagger}$) and reaction energies ($\Delta$G$^{rx}$) for the different H transfers to finally form \hts. Units are in \kcalMol. }
            \begin{tabular}{c| c c c}
                \hline
                \hline
                  & & \hydrogen-Penta-Fe & \hydrogen-Tetra-Fe \\
                  & System & \begin{tabular}{c c} TS FeSH-H & TS FeSH$_{2}$ \end{tabular} & \begin{tabular}{c c} TS FeSH-H & TS FeSH$_{2}$ \end{tabular} \\
                  \hline
                 $\Delta$G${^\ddagger}$ & \begin{tabular}{c}
                      Tetra-Def \\
                      Penta-Def \\
                      Tetra-Def-2 \\
                      Penta-Def-2 \\
                      Tetra-Penta-Def 
                 \end{tabular} 
                    & \begin{tabular}{c c}
                        22.7 & 33.3 \\
                        24.0 & 21.2 \\
                        18.1 & 26.5 \\
                        18.6 & 20.1 \\
                        16.1 & 20.7
                    \end{tabular} 
                    & \begin{tabular}{c c}
                        21.7 & 55.0 \\
                        21.9 & 27.3 \\
                        17.2 & 48.0 \\
                        0.1 & 37.5 \\
                        7.6 & 36.5
                    \end{tabular}  \\
                 \hline
                 $\Delta$G$^{rx}$ & \begin{tabular}{c}
                      Tetra-Def \\
                      Penta-Def \\
                      Tetra-Def-2 \\
                      Penta-Def-2 \\
                      Tetra-Penta-Def 
                 \end{tabular} 
                    & \begin{tabular}{c c}
                        10.7 & 29.5 \\
                        -3.7 & 20.1 \\
                        12.5 & 22.5 \\
                        5.1 & 10.6 \\
                        -1.1 & 13.8
                    \end{tabular} 
                    & \begin{tabular}{c c}
                        11.7 & 45.0 \\
                        2.59 & 29.4 \\
                        11.4 & 38.5 \\
                        -12.0 & 24.8 \\
                        -1.3 & 21.8
                    \end{tabular}  \\
                \hline
            \end{tabular}
            \label{tab:GibbsE}
        \end{table*}

    \subsection{\hts\ desorption} \label{sec:hts_desorption}

        Finally, we study the desorption of \hts\ from the surface. Despite the differences amongst the different systems, desorption energies are always above 40 \kcalMol. The different energetic values between the systems can be attributed to the different geometrical reorganisation after \hts\ desorption. In all cases, breaking the \hts-Fe bond is highly energy demanding, and the final surface becomes highly unstable. This value is in agreement with the experimental value found in this work, and points to the fact that the experimental $E_{a}$ estimate reflects the desorption energy, and not the reaction energy barriers. Hence, the desorption can be understood as the rate determinant step of the mechanism.
        
        We can compare our desorption energy result to binding energies of \hts, which have been studied in a different set of surfaces, both experimentally and theoretically. Temperature Programmed Desorption (TPD) experiments of \hts\ in water ice show a binding energy of $\approx$ 5 \kcalMol\ \citep{Collings2004, Jimenez-Escobar2011, Cazaux2022, Furuya2022}. While comparison with theoretical values is difficult, due to the pre-exponential factors used to derive them \citep{Ferrero2022, Minissale2022}, experimental values agree with binding energies calculated through quantum mechanical computations, which range from 1.5 to 8.5 \kcalMol\ \citep{Oba2018, Perrero2022, Ferrero2022, Perrero2024, Bariosco2024}. As it has been discussed in the literature, these low values explain why \hts\ is not found in icy grain mantles, and discards \hts\ as a S reservoir on ices. Binding energies of \hts\ in silicates have also been studied \citep{Perrero2024}. Their work found that adsorption of \hts\ in silicates is more favourable due to the dative covalent interactions between S and Mg or Fe, compared to the H-bond interactions present on ices. These covalent interactions are more stable for the S-Fe (~23 \kcalMol), while the S-Mg is less stable (17-20 \kcalMol) and promotes dissociative adsorption. Their results argue that already formed \hts\ present in the ISM is likely to stick to grain surfaces and become part of their cores as refractory material.

        The desorption energy measured in this work is much larger ($\geq$ 40 \kcalMol) due to the large destabilisation of the \fes\ surface upon \hts\ desorption, compared to the silicate surfaces. Our results support the hypothesis that once in the refractory phase of dust grains, S is hard to remove at interstellar conditions, needing of very energetic events to bring it back to the gas phase. This opens the door to explain both the depletion of S from the gas phase during the first stages of star formation and the lack of detection in the icy dust grains mantles during these stages, as an important amount of S might be locked in the refractory phase of dust grains.

%%%%%%%%%%%%%%%%%%%%%%%%%%%%%%%%%%%%%%%%
%	Conclusions   %
%%%%%%%%%%%%%%%%%%%%%%%%%%%%%%%%%%%%%%%%

\section{Conclusions} \label{sec:conclusion}

    In view of the results, we conclude that the formation of \hts\ on the FeS surface through subsequent hydrogenations is feasible at high temperatures. While the most energetic step is the desorption of \hts\ from the surface ($\geq$ 40 \kcalMol), at the working temperatures \hts\ is easily released to the gas phase. We can prove that at the experimental conditions, the process is favourable and possible, however, its importance in different environments of the ISM needs to be further discussed.

    Hydrogen sulphide has been identified through the years in different astronomical sources, such as molecular clouds \citep{Thaddeus1972}, star forming regions \citep{Holdship2016}, in cold regions of protoplanetary disks \citep{Phuong2018}, and the atmospheres of some Solar System planets \citep{Irwin2018, Irwin2019}. Since the temperature in these regions is really low, between 10-150 K, this reaction would not likely produce significant quantities of \hts, nor allow from its desorption, and the S would remain in the solid phase, in the form of refractory \fes. At these conditions, formation of \hts\ might proceed instead through direct hydrogenation of S atoms that might have been sputtered in the surface of dust grains.

    While at interstellar conditions, S would remain in the refractory phase, we can envision a likely environment for the formation and release of \hts in the surface of recently formed planets. The first Earth atmosphere was probably formed by the out-gassing of material similar to chondritic material at elevated temperatures ($>$ 700 K), and mainly composed of \co, \cotw, \hydrogen, \nt\ and \water. Models show that this material might also out-gas \hts\ at temperatures larger than 500 K after accreting onto the planet \citep{Schaefer2010}. Our results show that at the working temperatures there is no out-gassing of \hts\ from the solid, suggesting that this process starts at a larger temperatures. In protoplanetary conditions, the proposed mechanism might add important quantities of \hts\ in early atmospheres at relatively low temperatures, compared to \hts\ out-gassing. The formed \hts\ could in this conditions proceed to form other complex molecules, such as thiols \citep{Llorca2000}. Interestingly, \hts\ is proposed to be intimately linked with the origin of life for its versatility in forming other compounds and as a possible source of energy, which remains present in living systems up to these days \citep{Olson2016}.

    In summary, our results support previous results which suggest that S is depleted from the gas phase and onto the refractory phase of dust grains. Moreover, they point towards an interesting route towards the formation of \hts\ and the simple removal of S from the refractory phase to a more reactive species, capable of produce further chemistry. While the proposed reaction is only favoured in high temperature environments, it leads to important products, i.e., \hts, which can be involved in further reactions with components present on the surface of dust grains or in the primeval Earth environment.

\section*{Acknowledgements}

    The Heriot-Watt University group acknowledges funding from the UK Research and Innovation and the Engineering and Physical Sciences Research Council (UKRI-EPSRC) for the project \textit{Astrocatalysis: In Operando Studies Of Catalysis And Photocatalysis of Space-abundant Transition Metals}, grant number EP-W023024-1. The group also acknowledges the services of the  Mechanical Engineering Workshop of the school of Engineering and Physical Sciences (EPS), and the use of the High Performance Computer (HPC) at the Computational Engineering Research (CER) group at the Institute of Mechanical, Process and Energy Engineering (IMPEE).

    The Universitat Aut\`onoma de Barcelona group acknowledges funding within the European Union’s Horizon 2020 research and innovation program from the European Research Council (ERC) for the project \textit{Quantum Chemistry on Interstellar Grains} (QUANTUMGRAIN), grant agreement No 865657, and the funding within the the Marie Skłodowska-Curie Actions (MSCA) Postdoctoral Fellowships program from the European Commission for the project \textit{Computational Studies on Heterogeneous Astrocatalysis of Space-Abundant Transition Metals} (CHAOS), grant agreement HEU-101105235-CHAOS. G.P. thankfully acknowledges financial support by the Spanish Ministry of Universities and the European Union’s Next Generation EU fund for a Margarita Salas contract. Spanish MICINN is also acknowledged for funding the projects PID2021-126427NB-I00 and CNS2023-144902. The authors thankfully acknowledge RES resources provided by Univ. Valencia for the use of Tirant (activity QHS-2022-2-0022) and by BSC for the use of MareNostrum (activity QHS-2023-1-0019). The supercomputational facilities provided by CSUC are also acknowledged. The EuroHPC Joint Undertaking through the Regular Access call project no. 2023R01-112, hosted by the Ministry of Education, Youth and Sports of the Czech Republic through the e-INFRA CZ (ID: 90254) is also acknowledged.

    J.L. is a Serra H\'unter Fellow and is grateful to the Instituci\'o Catalana de Recerca i Estudis Avan\c{c}ats (ICREA) Academia program and project GC 2021 SGR 01061.

%%%%%%%%%%%%%%%%%%%%%%%%%%%%%%%%%%%%%%%%%%%%%%%%%%
\section*{Data Availability}

 Supporting material for the computational analysis, including Cartesian coordinates of the optimised minima and TS, the vibrational analysis and examples of input and output files can be found in the associated Zenodo repository (\url{https://doi.org/10.5281/zenodo.13322595}).



\bibliographystyle{mnras}
\bibliography{example} % if your bibtex file is called example.bib

%\begin{thebibliography}{99}
%\end{thebibliography}

%%%%%%%%%%%%%%%%%%%%%%%%%%%%%%%%%%%%%%%%%%%%%%%%%%

%%%%%%%%%%%%%%%%% APPENDICES %%%%%%%%%%%%%%%%%%%%%

% \appendix

% \section{Some extra material}

%%%%%%%%%%%%%%%%%%%%%%%%%%%%%%%%%%%%%%%%%%%%%%%%%%

% Don't change these lines
\bsp	% typesetting comment
\label{lastpage}
\end{document}

\maketitle

\begin{sloppypar}

\horrule{1pt} \\ [0.4cm]

\section{Complete summary of the values for the reaction PES} \label{ap:compu_summary}

    In this Section, we present the complete results regarding the Gibbs relative energies at 800 K, for all the steps regarding the formation of \hts, shown in Fig 3. All the obtained values are shown in Table \ref{tab:gibbsE_full}, where the results are divided first considering the type of defect, and the position where \hydrogen\ has been absorbed. 

    \begin{sidewaystable}
        \centering
        \begin{tabular}{c | c c c c c c c}
            \hline
            \hline
            System & FeS & FeS-H$_{2}$ & TS FeSH-H & FeSH-H & TS FeSH$_{2}$ & FeSH$_{2}$ & Fe + H$_{2}$S \\
            \hline
            \begin{tabular}{c|c}
                Tetra-Def & H$_{2}$-Penta-Fe \\ & H$_{2}$-Tetra-Fe \\
                Penta-Def & H$_{2}$-Penta-Fe \\ & H$_{2}$-Tetra-Fe \\
                Tetra-Def-2 & H$_{2}$-Penta-Fe \\ & H$_{2}$-Tetra-Fe \\
                Penta-Def-2 & H$_{2}$-Penta-Fe \\ & H$_{2}$-Tetra-Fe \\
                Tetra-Penta-Def & H$_{2}$-Penta-Fe \\ & H$_{2}$-Tetra-Fe 
            \end{tabular}
            &
            \begin{tabular}{c} 0.0 \\ 0.0 \\ 0.0 \\ 0.0 \\ 0.0 \\ 0.0 \\ 0.0 \\ 0.0 \\ 0.0 \\ 0.0 \end{tabular} 
            &
            \begin{tabular}{c} 1.9 \\ -12.6 \\ 0.7 \\ -20.1 \\ 2.0 \\ -17.1\\ -1.1 \\ -14.2 \\ -6.9 \\ -20.8 \end{tabular} 
            &
            \begin{tabular}{c} 22.7 \\ 9.1 \\ 24.0 \\ 1.8 \\ 18.1 \\ 0.1 \\ 17.6 \\ -14.1 \\ 9.9 \\ -13.3 \end{tabular} 
            &
            \begin{tabular}{c} 10.7 \\ -0.9 \\ -3.7 \\ -17.5 \\ 12.5 \\ -5.7 \\ 4.0 \\ -26.2 \\ -7.2 \\ -22.2 \end{tabular} 
            &
            \begin{tabular}{c} 33.2 \\ 42.4 \\ 17.5 \\ 9.8 \\ 26.5 \\ 30.9 \\ 19.0 \\ 11.3 \\ 13.4 \\ 14.3 \end{tabular} 
            &
            \begin{tabular}{c} 29.5 \\ 32.3 \\ 16.4 \\ 11.9 \\ 22.5 \\ 21.4 \\ 9.5 \\ -1.4 \\ 6.6 \\ -0.4 \end{tabular} 
            &
            \begin{tabular}{c} 48.0 \\ 54.5 \\ 57.5 \\ 62.2 \\ 49.2 \\ 50.1 \\ 55.7 \\ 39.3 \\ 50.6 \\ 39.7 \end{tabular}

        \end{tabular}
        \caption{Collected Gibbs relative energies (T = 800 K) in \kcalMol, for the formation of \hts, including the \hydrogen\ adsorption and \hts\ desorption. The different proposed mechanisms are divided first considering the type of defect, and secondly, the position where \hydrogen\ has been adsorbed}
        \label{tab:gibbsE_full}
    \end{sidewaystable}

\end{sloppypar}